\documentclass[conference,a4paper]{IEEEtran}

\usepackage{amsmath,amssymb}
\usepackage{array}
\usepackage{booktabs}
\usepackage{tabularx}
\usepackage{url}
\usepackage{cite}
\usepackage{balance}
\usepackage[T1]{fontenc}
\usepackage{times}

\newcolumntype{Y}{>{\centering\arraybackslash}X}
\newcolumntype{L}{>{\raggedright\arraybackslash}X}

\title{Trust, but Validate the Instrument: Auditing\\
AI-Generated RTL Verification Plans on Authored\\
Security-Regression Proxies}

\author{%
\begin{tabular}{ccc}
\begin{tabular}[t]{c}
Hang Xiao\\
Fortinet, Inc.\\
Sunnyvale, CA, USA\\
xhang@fortinet.com
\end{tabular}
&
\begin{tabular}[t]{c}
Chuhong Xu\\
Sony Corporate of America\\
San Jose, CA, USA\\
chuhong.xu@sony.com
\end{tabular}
&
\begin{tabular}[t]{c}
Kainan Zhou\\
Google LLC\\
Mountain View, CA, USA\\
zhoumark@google.com
\end{tabular}
\\[1.5em]
\multicolumn{3}{c}{%
\begin{tabular}{cc}
\begin{tabular}[t]{c}
Gangzhen Qian\\
Google LLC\\
Mountain View, CA, USA\\
irisqian@google.com
\end{tabular}
& \hspace{3em}
\begin{tabular}[t]{c}
Lu Yi\\
Google LLC\\
Mountain View, CA, USA\\
annabelyi@google.com
\end{tabular}
\end{tabular}}
\end{tabular}%
}

\begin{document}
\maketitle
\vspace*{2.5pt}

\begin{abstract}
AI-generated RTL verification plans can satisfy a provider schema yet fail at the boundary to trusted execution. We present SecTB-RTL, an auditable framework covering 31 tasks and 124 authored hardware-security regressions. A deterministic non-AI baseline killed 36, 75, and 78 mutants at increasing resource limits. The first confirmatory run (C1-R2) failed before model execution because the provider rejected its response schema. After a schema-only repair made without viewing outcomes, a separately frozen follow-up run (C1-R3) completed 1,860 calls. The provider accepted 1,857 responses, but only nine passed the production semantic validator. The generation and execution rules did not match. We therefore preserve the run as an instrument-validation incident and report no prompt-effect estimate. This incident shows that provider or schema acceptance does not establish execution validity. Compilation and coverage are only diagnostics; the exact saved artifact must pass the full production path. A subsequent follow-up is excluded because it did not satisfy the preregistered evidence-completeness gate and is treated only as future work. We release the benchmark, failure-preserving contract, incident provenance, and governance controls needed to prevent infrastructure behavior from being misreported as model behavior.
\end{abstract}

\vspace{10pt}

\section{Introduction}
AI-assisted chip design is moving from code completion toward the generation of verification artifacts. AutoBench demonstrates that LLMs can generate self-checking HDL testbenches from a design-under-test description \cite{qiu2024autobench}. This creates a trust-calibration problem. An artifact may parse, run, and cover much of the design yet lack the stimulus or oracle needed to detect a security-property violation. Compilation, a passing golden run, and high line coverage are therefore insufficient on their own.

This paper studies that problem at the intersection of cybersecurity, AI applications, and RTL/ASIC design. Here, the AI system generates security-verification stimuli and checks. The goal is to detect hidden CWE-specific regressions in synthesizable RTL evaluated with standard simulation, synthesis, and formal tools. The resulting question is not whether LLMs can emit HDL. It is whether generated verifiers provide measurable security evidence under a protocol that counts every scheduled attempt, including failures.

Prior testbench studies primarily evaluate functional correctness and executable behavior. Security-oriented hardware benchmarks show why functionally acceptable generated RTL can still carry CWE-class weaknesses \cite{chen2026hardsecbench}. These advances do not answer three narrower deployment questions. First, how much apparent success is lost between a model response and a verifier that is valid on the golden design? Second, does security context improve detection relative to a functional prompt with equal attention and output budget? Third, when a generated verifier reaches high structural coverage, how often can a qualified security regression remain undetected by the generated oracle?

We use task-level studies whose prompts, schedules, and scoring rules were fixed before provider contact, rather than a collection of successful examples. The scientific unit in the planned treatment experiment is the RTL task, not an individual mutant run. Each artifact is tested unchanged on a known-correct design and four hidden mutants. Every upstream failure remains among the scheduled attempts. The treatment experiment did not reach an analyzable endpoint. Instead, the incident shows how the measurement process can fail before an effect can be estimated.

\subsection{Research questions}
\begin{itemize}
\item RQ1---Instrument attrition: How much provider-accepted output survives the exact production semantic boundary?
\item RQ2---Assurance boundary: Which production-semantic failures remain invisible to provider or schema acceptance? Can a fail-closed lifecycle, which blocks claims after failed calibration, prevent them from becoming scientific claims?
\item RQ3---Attainable reference: What mutant-detection yield is achieved by a deterministic, outcome-blind non-AI stimulus baseline at fixed resource limits?
\end{itemize}

\subsection{Contributions}
This paper makes four bounded contributions.
\begin{enumerate}
\item \textit{A security-evidence benchmark.} SecTB-RTL's amended confirmatory frame contains 31 RTL tasks in 16 (CWE, root mechanism) groups, with four hidden, prequalified regressions per task. The 124 mutants retained the same content and qualification status across the R2 rebuild and the separately frozen R3 incident study.
\item \textit{End-to-end incident evidence.} A complete 1,860-call run quantifies the gap between provider acceptance (1,857 calls) and production-semantic validity (nine calls). The incident is preserved in full rather than restricted to nine valid calls or presented as a treatment result.
\item \textit{A fail-closed assurance contract.} ``Fail closed'' means that failed checks do not count as valid evidence and failed calibration blocks treatment claims. Separate gates cover provider acceptance, schema, production semantics, rendering, two-simulator agreement, golden validity, and mutant detection. Cryptographic hashes identify the exact frozen files; limited amendment rules prevent post-result retries or silent repairs.
\item \textit{Calibrated reference evidence.} The deterministic baseline reports yield at three budgets on all 124 mutants, showing benchmark feasibility without implying an AI prompt effect.
\end{enumerate}

The paper does not claim the first LLM-generated hardware-security verification plan, universal improvement from security prompting, exploit prevention, or security preservation through synthesis, physical implementation, or field deployment.

\section{Background and Related Work}
Prior work establishes that LLMs can produce several kinds of hardware-verification artifact, but the evaluation target varies. Open benchmarks assess generated UVM-oriented testbenches \cite{murthy2025verifllm} and candidate assertions \cite{pulavarthi2025assertionbench}. SecTB-RTL asks a different question: whether a generated plan and its trusted execution chain together form a valid security-measurement instrument.

Our contribution is therefore methodological rather than priority-based. It combines task-level treatment assignment, hidden regressions qualified in advance, and a deployment process that rejects evidence after a failed check. Mutation testing supplies controlled faulty variants; each hardware-security variant must also weaken an invariant, be reachable under legal inputs, and affect a top-level output. Likewise, structural coverage can diagnose stimulus reach without proving that the generated artifact contains the correct security oracle. SecTB-RTL therefore measures evidence that survives a workflow fixed in advance, rather than an occasionally useful test or assertion.

This framing connects AI-assisted hardware vulnerability assessment with trustworthy deployment. The output is a benchmark, scoring pipeline, and evidence about when users should trust generated verification artifacts---not a claim to the first LLM security testbench or a general EDA coverage result.

\section{Security and Evaluation Model}
\subsection{Defensive use case}
We consider a verification engineer who possesses synthesizable golden RTL and a functional specification, and asks an AI model for a deterministic stimulus-and-oracle plan in a restricted JSON DSL. A trusted harness validates the plan and lowers it to a self-checking SystemVerilog testbench. Depending on the experiment arm, the engineer may also provide a neutral checklist, an abstract threat model, or explicit security requirements. The generated plan is intended to screen future RTL changes for regressions through top-level behavior. Prior work shows that language models can synthesize hardware-security assertions \cite{kande2024security}; our unit is instead a persisted, executable stimulus-and-oracle plan that must traverse the trusted production path unchanged.

\subsection{Assets, attacker capability, and violations}
Each task defines an asset, a set of legal attacker-controlled top-level inputs, and an invariant. Representative root mechanisms include authorization/gating, secure reset or fail-safe defaults, sticky and priority state, boundary/comparison checks, and leakage through debug or scan behavior. A CWE taxonomy supplies the vulnerability-class vocabulary \cite{mitre2026cwe}. A mutant is a controlled regression that violates the invariant for at least one legal trace while preserving the module interface and ordinary execution environment used across the benchmark.

The evaluation adversary is not assumed to compromise the model service or the scoring host. Instead, it represents an RTL change---accidental or malicious---that weakens a security property. Detection requires both an activating stimulus and an oracle that rejects the resulting top-level behavior. Merely reaching the changed logic is insufficient.

\subsection{Information boundary}
Every model arm receives the same alpha-renamed, comment-free golden RTL, functional requirements, port contract, output schema, token budget, and execution contract. Models never receive mutant RTL, diffs, witness traces, hidden official tests, scorer-only identifiers, or mutation outcomes. SA and SE reveal no CWE identifier or implementation defect. This boundary tests conditioning on legitimate design information, not direct localization of a known mutation.

\subsection{Non-goals}
The experiment does not model physical attacks, analog leakage, placed-and-routed behavior, gate-level timing, foundry threats, or post-silicon validation. It does not establish that a killed mutant corresponds to a practical exploit or that a surviving test implies an exploitable chip. It evaluates detection of controlled RTL security regressions under the top-level environment defined by the benchmark.

\section{SecTB-RTL Benchmark Construction}
\subsection{Source frame and stratification}
The construction frame is derived from HardSecBench revision \texttt{e2084be7} \cite{chen2026hardsecbench}. Tasks are purposively selected and screened, not probability-sampled from all RTL designs. The historical C1 frame contained 32 tasks in 16 two-task (CWE, root mechanism) strata. Before any provider response, the treatment audit excluded T21 because its golden RTL could not instantiate the intended construct and no outcome-blind reserve qualified. C1-R2 therefore retains 31 tasks and all 16 strata: 15 contain two tasks, while CWE-1241 contains one. Seven retained tasks (22.6\%), spanning seven strata, keep their original one-shot holdout labels. The primary analysis includes all 31 tasks; holdout-only and stratum-level results are descriptive. After source T21 is excluded, R2 runner slots T21--T31 project source slots T22--T32. Consequently, the prespecified high-ceiling source sensitivity T04/T13/T27 maps to runner T04/T13/T26; runner T27 is source T28 and is not part of that exclusion or the high-ceiling sensitivity set.

The task-selection rule, inclusion and exclusion reasons, alpha-renaming, and prompt materialization are fixed independently of formal C1 model outputs. Public benchmark membership and screening criteria are disclosed because they limit population and contamination claims.

\subsection{Security-regression mutants}
Four mutants are authored per task before model generation. Following mutation-based testbench qualification \cite{huang2015mutation}, each mutant weakens the same task-level security objective through a semantically applicable change but follows a distinct behavior or witness. Candidate operators cover omitted authorization, weakened reset/default behavior, non-sticky state, incorrect priority or transitions, boundary and comparator changes, and debug/scan leakage. Mutants do not change top-level ports.

Every included mutant must pass all of the following gates:
\begin{enumerate}
\item parse and compile in the frozen tool environment;
\item exact top-level interface identity with the golden RTL;
\item synthesis of both golden and mutant RTL;
\item a legal witness that activates and exposes the difference through the generated top-level harness ports;
\item formal non-equivalence to the golden design;
\item replay of the witness under the trusted harness; and
\item pairwise formal distinctness among the task's four mutants.
\end{enumerate}

Equivalent, unreachable, unobservable, interface-changing, or duplicate variants are excluded before any generated-verifier outcome exists for a study cell.

Table~\ref{tab:scale} separates benchmark qualification from generation-study scale. Historical qualification and retained-frame counts are benchmark evidence, not AI performance.

Qualification shows that the evaluation targets are executable, reachable, observable, and distinct. It does not show that any AI-generated verification plan will detect them.

\begin{table}[t]
\caption{Benchmark and closed R3 incident scale.}
\label{tab:scale}
\centering
\footnotesize
\setlength{\tabcolsep}{3pt}
\begin{tabularx}{\columnwidth}{@{}Lcc@{}}
\toprule
Item & Retained benchmark & Closed R3 incident \\
\midrule
RTL tasks & 31 & 31 \\
(CWE, mechanism) strata & 15 paired + 1 singleton & same \\
Included mutants & 124/124 qualified & 124 \\
Model variants / prompt arms & --- & 3 / 4 \\
Repetitions & --- & 5 \\
Complete randomized blocks & --- & 465 \\
Scheduled generations & --- & 1,860 \\
Scheduled mutant rows & --- & 7,440 \\
\bottomrule
\end{tabularx}
\end{table}

\section{Evaluation Contract}
Only 9 of 1,860 C1-R3 calls met the production executor's semantic rules, so we retain the run as an instrument-validation incident. This section describes the rules fixed before provider contact and why the completed call schedule cannot support a treatment-effect estimate.

Specification-driven systems can derive candidate verification properties from design documents \cite{yan2025assertllm}. In our study, the schema, token budget, and base prompt are fixed. Only the supplied information varies. F0 gives the functional objective. FC adds an equal-attention functional control. SA adds abstract security context without CWE or mutant details. SE states explicit security requirements. The primary contrast is SE-FC; secondary contrasts are SA-FC and FC-F0.

\subsection{Matrix and randomization}
R3 crosses 31 tasks, three model aliases, four prompt arms, and five repetitions. It yields 465 complete task-model-repetition blocks, 1,860 generations, and 7,440 mutant rows. Arm assignment is randomized within each complete block, and the global invocation order is randomized independently of treatment. Each cell allows one provider attempt. Calls are not retried, and assignments are not redrawn. Every scheduled cell remains in the denominator, including provider rejections, malformed outputs, compilation failures, simulator mismatches, and other terminal failures.

\subsection{Generated-verifier contract}
Prior plan-generation work separates synthesis from execution and evaluates golden validity alongside mutation detection \cite{kochar2026grpo}. SecTB-RTL records this boundary explicitly. Every R3 call had to return \texttt{sectb-stimulus-v1} JSON with \texttt{dsl\_version}, \texttt{test\_name}, and 4--128 ordered steps. The DSL permits reset assignment, top-level driving, bounded ticks (1--16), and output assertions. It prohibits random sources, loops or branches, hierarchy access, force, DPI, file I/O, and model-authored verdicts. The generation validator checked only whether responses followed this JSON language. The execution layer separately checked task interfaces and reset behavior, then rendered and ran passing responses in Icarus and Verilator. A separate three-call probe tested schema acceptance and the generation validator, but not reset rules or the full simulator path.

\subsection{End-to-end outcome}
For task $t$, model $j$, arm $a$, repetition $r$, and mutant $m$:
$K_{tjarm}=1$ only when the response satisfies the fixed response rules, passes golden validation in both simulators, and kills the mutant; otherwise $K_{tjarm}=0$.
\begin{align}
\bar{Y}_{t,a} &= \frac{1}{60}\sum_{j=1}^{3}\sum_{r=1}^{5}\sum_{m=1}^{4} K_{tjarm}, \label{eq:ybar}\\
\theta(a) &= \frac{1}{31}\sum_{t=1}^{31}\bar{Y}_{t,a}, \label{eq:theta}\\
\Delta &= \theta(\mathrm{SE})-\theta(\mathrm{FC}). \label{eq:delta}
\end{align}
Failed calls remain in the analysis; we do not select only cases that reach a later stage.

\subsection{Statistical analysis}
The registered analysis uses the task as the statistical unit ($n=31$); model, repetition, and mutant observations are nested within task. It specifies paired contrasts, a bootstrap that resamples whole tasks, and randomization within each block. The primary test is SE-FC at $\alpha=0.05$; secondaries are Holm-adjusted. No hierarchical confirmatory model is used. R3 did not pass calibration against the production semantic rules, so these procedures are not used for confirmatory prompt-effect claims.

\subsection{Coverage blind spots and failure decomposition}
A coverage blind spot is a mutant that survives conclusive execution even though the verifier passes the golden design and meets the coverage thresholds (line $\ge 90\%$, toggle $\ge 80\%$). The funnel is:
\begin{center}
\footnotesize
scheduled $\rightarrow$ response/schema $\rightarrow$ compile $\rightarrow$ golden-valid\\
$\rightarrow$ conclusive execution $\rightarrow$ kill
\end{center}
Diagnostics record golden false positives, simulator agreement, task/model breakdowns, receipt integrity, latency, tokens, and cost. They cannot support a treatment estimate when generation and execution enforce different rules.

\section{Preregistration, Implementation, and Reproducibility}
Before provider contact, we fixed the task registry, mutants, qualification evidence, prompts, schema, schedule, scorer, and analysis rules. The historical release is tagged \texttt{c1-confirmatory-author-freeze-v1} at commit \texttt{2db72f31}. A manifest records a cryptographic hash for every eligible file; each call checks that the working tree is unchanged. The public repository retains the tagged lineage for independent inspection, although the preregistration was author-sealed rather than independently reviewed.

The first operational C1 namespace failed before a provider response because the desktop sandbox could not write the local state database. Its record contains exit code 1, empty provider output, no request identifier or provider event, and no model response. We retain it as a pre-provider infrastructure incident, not model behavior, and exclude it from every confirmatory denominator reported here.

Before any later research call, an outcome-blind audit (performed without viewing treatment outcomes) found functional-specification conflicts and security-label leakage. We therefore stopped C1-R1 before its first research call. The amended C1-R2 excluded T21, retained 31 tasks and 124 mutants, and changed only documented COMMON/SA/SE text. All tasks passed checks C1--C6, and 31/31 passed the C7 byte-identity check across arms. We rebuilt the related registries and qualification records without changing scientific content. A second audit found that the unused schedule randomized labelled arm order rather than assignment. Before provider contact, we preserved it, drew one 256-bit seed without redraw, and used one of the $4!$ possible arm assignments for each of 465 blocks; invocation order was randomized separately. The R2 availability probe tested service access, not the research schema. These changes were documented, checked, and sealed at commit \texttt{675c214} under tag \texttt{82cecde5}; they are amendments, not part of the original preregistration.

R2's first eight scheduled calls were rejected before model execution because fixed op values lacked explicit JSON types. The immutable records contain one attempt each and no model response, token, cost, execution, coverage, or security outcome. R2 is therefore a closed schema incident. C1-R3 was separately frozen, not a retry. It reused the 465 assignments and 1,860-call order without redraw. The repair added types to singleton enums, replaced \texttt{oneOf} with \texttt{anyOf}, and removed two \texttt{uniqueItems} keywords. Prompts, RTL, contracts, arms, and the production validator otherwise remained fixed; the schema probe, updated evidence links, and author freeze preceded generation.

As Section~VII reports, R3 completed the schedule but failed the production-semantic calibration check. An outcome-blind replay showed that the provider-visible rules and probe did not enforce the executor's reset state machine; even the canonical probe response failed. We retain the complete 7,440-row denominator and funnel, but neither analysis restricted to valid cases nor an arm contrast. Counting failures as zero is valid only when generation and execution rules are aligned; it cannot repair this mismatch.

The frozen implementation uses Icarus Verilog and Verilator for cross-simulator execution and Yosys for synthesis and formal preparation. Artifact manifests record executable hashes, software versions, unavailable provider controls, and the code for artifact creation, scoring, and analysis. Incident artifacts remain separate from confirmatory evidence.

\begin{table*}[t]
\caption{Evidence retained and scientific disposition.}
\label{tab:evidence}
\centering
\footnotesize
\setlength{\tabcolsep}{4pt}
\begin{tabularx}{\textwidth}{@{}p{0.12\textwidth}p{0.18\textwidth}L p{0.27\textwidth}@{}}
\toprule
Study & Evidence retained & Result & Scientific disposition \\
\midrule
S1-R2 deterministic baseline & 124 fixed mutant rows at each budget & 36/124, 75/124, and 78/124 kills at 32, 64, and 128 steps; task-macro yields 0.290, 0.605, and 0.629 & Valid bounded comparator; not an exchangeable AI arm \\
C1-R2 & Eight provider attempts & All rejected at provider schema prevalidation; no model responses & Closed schema incident \\
C1-R3 & 1,860 completed calls & 1,857 provider-accepted, three provider failures, nine production-semantic-valid & Closed instrument-validation incident; no effect estimate \\
\bottomrule
\end{tabularx}
\end{table*}

\section{Results}
The available evidence answers instrument and calibration questions only. Table~\ref{tab:evidence} separates benchmark calibration from the two closed AI-study incidents.

\subsection{R2 failed before response generation; R3 failed at production semantics}
C1-R2 stopped after eight provider attempts because the frozen response schema was rejected during provider prevalidation. No model response, executable artifact, or security outcome exists for that run.

C1-R3 completed the frozen schedule, but production-semantic validity was 9/1,860 (0.48\%) despite provider transport acceptance of 1,857/1,860 (99.84\%). The 1,848-call gap was traced, before arm-, model-, kill-, or effect-stratified inspection, to a reset-state-machine rule enforced by execution but absent from the generation-facing contract and exact-schema probe. The probe's own canonical response failed the later production rule. Failure-as-zero scoring is valid for failures under a calibrated instrument; here it would chiefly quantify an undisclosed instrument mismatch. We therefore report the funnel but no treatment contrast or complete-case rescue of treatment effects.

\subsection{The deterministic baseline calibrated attainable yield}
The non-AI generator killed 36/124 mutants at 32 steps, 75/124 at 64 steps, and 78/124 at 128 steps. The corresponding task-macro yields were 0.290, 0.605, and 0.629. The gain from 64 to 128 steps was small relative to the first budget increase, indicating that longer generic stimulation alone does not remove the need for security-specific activation and oracles. Because the baseline uses a fixed hand-authored template, it is a benchmark calibration rather than an exchangeable arm in the registered treatment design.

\subsection{Failure boundaries and governance controls}
Table~\ref{tab:boundaries} summarizes the operational lesson from R1--R3: each boundary requires its own detector, and a later successful stage cannot retroactively validate an earlier mismatch.

\section{Discussion}
C1-R3 quantifies a large and operationally important gap: structured output and provider acceptance did not imply compatibility with the production verifier. The incident motivates a layered assurance chain that must be calibrated before research execution through the following sequence:
\begin{center}
\footnotesize
provider $\rightarrow$ schema $\rightarrow$ semantic contract $\rightarrow$ renderer $\rightarrow$ persistence $\rightarrow$\\
resume replay $\rightarrow$ golden validation $\rightarrow$ mutant oracle
\end{center}
Each arrow is a potential semantic boundary. A probe that terminates before the next boundary cannot validate it. Canonical encoding, negative fixtures, and a complete create-only archive are therefore needed at production and replay.

The fail-closed lifecycle changed what could responsibly be claimed. R3's complete provider schedule was not mined for a favorable arm, and the incomplete follow-up is not promoted into an effect estimate. This costs a treatment result but prevents infrastructure behavior from masquerading as model behavior. The deterministic baseline remains useful because it calibrates the 124 mutants independently of the failed AI instrument, although its hand-authored oracle and resource budget make it a reference rather than an exchangeable competitor.

For practice, AI-generated verification assets should enter security decision making only after the exact persisted artifact---not an in-memory precursor---has passed production validation, clean golden execution, hidden-regression tests, and reproducible resume checks. Coverage-guided stimulus generation can use iterative feedback to exercise more behavior \cite{zhang2025llm4dv}, but compilation and coverage remain diagnostics; neither establishes an adequate oracle for a security claim.

\section{Threats to Validity}
\begin{itemize}
\item \textit{Construct validity:} Authored mutants are controlled regression proxies, not production exploits. R3's contract mismatch invalidates treatment interpretation even though its aggregate funnel is reliable incident evidence.
\item \textit{Internal validity:} Incident classification was made without arm-, model-, kill-, or effect-stratified inspection. No complete-case subset or repaired post hoc treatment contrast is reported.
\item \textit{Statistical validity:} R3 supports an aggregate incident funnel, not a prompt-effect estimate. The task-level analysis cannot repair construct invalidity in the measurement instrument.
\item \textit{External validity:} Evidence is bounded to public RTL, authored regressions, one provider surface, three aliases, fixed dates, and a restricted DSL. The result is a workflow-assurance finding, not a universal ranking of models.
\item \textit{Reproducibility:} Code, protocols, tags, and the R1--R3 lineage are versioned. Incident manifests are retained rather than overwritten or converted into treatment evidence.
\end{itemize}

\enlargethispage{3\baselineskip}
\section{Future Work}
A future confirmatory phase will retain the same 31-task benchmark and four-arm estimand while using a fresh study identity, an exact production-path probe, one attempt per scheduled cell, and a create-only evidence archive. Its arm, model, coverage, and mutant-kill statistics will be reported only after a complete denominator and its create-only archive are independently verified. This future phase is not part of the empirical evidence evaluated in the present paper.

\clearpage
\twocolumn[
\begin{minipage}{\textwidth}
\refstepcounter{table}\label{tab:boundaries}
\begin{center}
\footnotesize
TABLE \thetable\\
\textsc{Failure boundaries, detection methods, and governance measures.}
\end{center}
\vspace{-0.5em}
\centering
\footnotesize
\setlength{\tabcolsep}{4pt}
\begin{tabularx}{\textwidth}{@{}p{0.20\textwidth}p{0.32\textwidth}L@{}}
\toprule
Failure boundary & Detection method & Governance measure \\
\midrule
Local runner $\rightarrow$ provider process & Exit code, raw stdout/stderr, provider-event count & Preserve a terminal incident record; never infer a model response from an invocation failure \\
Provider $\rightarrow$ response schema & Provider prevalidation and strict duplicate-key JSON parsing & Freeze the exact provider-facing schema; use a separately excluded compatibility probe before research calls \\
Schema $\rightarrow$ production semantics & Replay the provider response through the exact production validator & Bind the same validator into prompt, probe, runner, and executor; stop the study when calibration fails \\
Semantic object $\rightarrow$ rendered harness & Byte-exact receipt and deterministic renderer replay & Persist response, receipt, and harness together; prohibit silent repair or post-response rewriting \\
Harness $\rightarrow$ executable evidence & Golden run, dual-simulator agreement, watchdog, and interface checks & Fail closed on infrastructure or harness disagreement and retain the scheduled denominator \\
Executions $\rightarrow$ scientific claim & Complete manifest, fixed denominator, task-level analysis, and provenance check & Require evidence completeness before inference; forbid complete-case rescue, retry, redraw, or selective reporting \\
\bottomrule
\end{tabularx}
\vspace{1em}
\end{minipage}
]

\section{Ethics and Responsible Release}
SecTB-RTL is intended for defensive evaluation of verification tools. It uses small, public or derived RTL modules and controlled regressions rather than confidential designs or deployable chip exploits. Released mutants demonstrate classes of authorization, state, boundary, and leakage errors and therefore have dual-use educational value; they do not include production keys, physical attack procedures, proprietary credentials, or real-world deployment targets.

Raw model outputs may contain unexpected text or malformed code and are treated as untrusted artifacts. Evaluation runs use isolated working directories, top-level-only HDL restrictions, bounded execution, and no generated file or network access. Credentials, private logs, and provider secrets are excluded from the artifact. Results will report failures and unknown metadata rather than silently discarding them, reducing pressure to overstate AI capability in a security-critical evaluation workflow or reporting process.

\section{Conclusion}
SecTB-RTL reframes AI-generated RTL verification-plan evaluation around the validity of the measurement chain. On 31 tasks and 124 authored regressions, a deterministic baseline establishes a bounded detection reference. The AI studies then expose two distinct failure layers: provider schema rejection in R2 and generation/execution semantic mismatch in R3. R3 completed 1,860 calls yet supports no prompt-effect claim under the frozen design and incident disposition.

The responsible result is therefore negative but concrete: provider acceptance, fluent structured output, and a schema-level probe did not validate a resumable security-evaluation instrument. Compilation and coverage remain downstream diagnostics, not substitutes for an aligned semantic contract. Trustworthy AI-assisted hardware verification must bind and test every boundary, including persistence and recovery, before outcome inspection. The result is a bounded account of where the measurement chain failed, how the failure was detected, and which controls prevent an unsupported model-performance claim.

\newpage
\section*{Artifact Availability}
The public repository contains the benchmark, frozen protocols, generation and execution code, qualification evidence, incident manifests, and provenance checks used here. The historical baseline freeze is tagged \texttt{c1-confirmatory-author-freeze-v1}. The repository retains C1-R2 as an immutable eight-attempt schema incident and C1-R3 as a completed 1,860-call instrument-validation incident. Their manifests, author freezes, exact-schema probe evidence, and superseded recovery records are preserved under distinct study namespaces. Neither incident is repackaged as a treatment result, and the subsequent incomplete follow-up is excluded from the paper's empirical evidence base reported here.

\section*{Acknowledgment}
ChatGPT(OpenAI) was used for language editing, experiment running, and assistance in script development. The authors reviewed the manuscripts, verified the reported results and citations, and take responsibility for the final text.

\end{document}